\documentclass[prb,twocolumn,showpacs,preprintnumbers,amsmath,amssymb]{revtex4}
\usepackage{graphicx}
\usepackage{dcolumn}
\usepackage{bm}

\begin{document}

\title{Electronic structure of
Ba(Fe,Ru)$_{2}$As$_{2}$ and Sr(Fe,Ir)$_2$As$_2$ alloys}

\author{Lijun Zhang}
\author{D.J. Singh}

\affiliation{Materials Science and Technology Division,
Oak Ridge National Laboratory, Oak Ridge, Tennessee 37831-6114}

\date{\today}

\begin{abstract}
The electronic structures of Ba(Fe,Ru)$_2$As$_2$ and Sr(Fe,Ir)$_2$As$_2$
are investigated using density functional calculations.
We find that these systems behave as coherent alloys from the
electronic structure point of view.
In particular, the isoelectronic substitution of Fe by Ru does
not provide doping, but rather suppresses the spin density wave
characteristic of the pure Fe compound by a reduction in the Stoner
enhancement and an increase in the band width due hybridization involving Ru.
The electronic structure near the Fermi level otherwise remains
quite similar to that of BaFe$_{2}$As$_{2}$.
The behavior of the Ir alloy is similar, except that in this case
there is additional electron doping.
\end{abstract}

\pacs{74.25.Jb,71.20.Be}

\maketitle

\section{introduction}

The discovery of unconventional superconductivity
in proximity to magnetism for layered Fe-based
materials\cite{kamihara08,delacruz,singh-du,mazin,dong,chen}
has led to considerable interest in both establishing the interplay
between magnetic order and superconducting state,
and searching for effective ways of tuning them.
The phase diagrams typically show a spin density wave (SDW)
that competes with superconductivity, i.e. superconductivity
generally appears when the SDW is suppressed either by doping with electrons
or holes, which reduces the nesting by making the electron and
hole Fermi surfaces mismatched or by pressure,
\cite{alireza}
which increases hybridization and broadens the bands again working against
nesting.

A remarkable feature of these compounds is that, in contrast to the cuprates,
they may be doped into a superconducting state
by substitutions on the Fe-site, e.g. in Ba(Fe,Co)$_2$As$_2$ and
Ba(Fe,Ni)$_2$As$_2$. \cite{sefat,ni-dope}
In these alloys the electronic structure remains very similar
to that of the parent Ba(Sr)Fe$_2$As$_2$, but the Fermi level is
shifted upwards corresponding to the increased electron count.
Thus one difference from cuprates is that substitution of Co and Ni
lead to the formation of a coherent alloy electronic structure rather
than the introduction of localized states associated with those ions.
Recently, it has been shown that in addition to the 3$d$ dopants, Co and
Ni, superconductivity can be induced by alloying with some 4$d$ and 5$d$
elements, including Ru, Rh, Ir and Pd.
\cite{ru_1,ru_2,ru_3,han-rh,sr_ir,zhu-pd}

Transition elements in the 4$d$ and 5$d$ rows differ from their
3$d$ counterparts in several respects.
Since $4d$ and $5d$ orbitals are much more extended than the
$3d$ orbitals there is a greater tendency towards covalency both
with ligands (e.g. As) and also in stronger metal -- metal $d$ bonds.
For example, mid-$5d$ transition elements have some of the highest
melting points of any material
(the melting points of Ir and Ru are 2739 K and
2607 K, respectively, as compared to 1811 K for Fe and 3695 K for W),
and compounds of these elements are
often extremely hard.
Thus one may expect broader bands and more hybridization with As in
the alloys with these elements. Secondly, again because of the
more extended $4d$ and $5d$ orbitals, the Hund's coupling on these
atoms is weaker than on $3d$ atoms, which works against magnetism and
is reflected in lower values of the Stoner parameter for $4d$ and $5d$
elements and compounds. Finally, the larger orbitals lead to a tendency 
for higher valence states in the $4d$ and $5d$ series, so that
Ru$^{4+}$ and Ru$^{5+}$ compounds are more stable and more common than
the corresponding Fe compounds.

Returning to superconductivity in alloys
of BaFe$_2$As$_2$ with $4d$ and $5d$ elements,
the case of Ru is particularly interesting
because Ru is isoelectronic with Fe and therefore it may or
may not serve as a dopant.
Transition temperatures
up to $T_{c}$ $\sim$ 21 K may obtained with substantial
nominal Ru content in Ba(Sr)Fe$_{2-x}$Ru$_{x}$As$_{2}$, $x_{\rm nom}\sim$0.7
although 
Ba(Sr)Ru$_{2}$As$_{2}$ shows neither magnetic order
nor superconductivity.\cite{baru2as2}
One possibility is that Ru serves as a dopant, supplying carriers
to the Fe planes. \cite{ru_1,ru_3}
For example, Ru might occur as Ru$^{4+}$, in which case empty localized
$d$ states associated with Ru atoms would occur above the Fermi energy,
while the Fe derived valence bands would show higher filling reflecting
electron doping by two carriers per Ru.
Another possibility is that alloy shows a coherent electronic structure
that is distorted from the pure Fe compound, but does not show additional
states,
similar to the Co and Ni doped materials.
In this case, the effect of Ru could be similar to that of pressure,
broadening the bands, increasing hybridization and/or lowering the
effective Hund's coupling and thereby destroying the SDW in favor
of superconductivity.

Here, we report density functional studies of the electronic structure
of BaFe$_{2-x}$Ru$_{x}$As$_{2}$.
It is found that the substitution of Fe with Ru gives a
quite similar electronic structure
to that of BaFe$_{2}$As$_{2}$
near the Fermi level $E_{F}$.
Importantly, this substitution
does not result in doping since we find no additional flat
bands reflecting Ru states in supercell calculations and consequently
the exact Luttinger's theorem compensation of electron and hole surfaces is
maintained in the same way as in the pure compound without the SDW.
Thus, Ru substitution does not
introduce additional electrons to this system.
Instead the SDW magnetic order is suppressed by the decreasing Stoner
term and increased hybridization.
We also performed the calculation for 5$d$ transition metal,
Ir alloyed SrFe$_{1-x}$Ir$_{x}$As$_{2}$,
which was reported to superconduct with $T_{c}$ $\sim$ 22 K.\cite{sr_ir}
We also find an electronic structure characteristic of a coherent alloy
for this non-isoelectronic, Co column transition metal.
We find electron doping as expected with additional one electron per Ir,
which cooperates with the reduction of Hund's coupling
to destabilize the SDW order.

\section{first principles calculations}

Our first principles electronic structure
calculations were performed within the
generalized gradient approximation (GGA) of
Perdew, Burke, and Ernzerhof (PBE),\cite{pbe}
using the general potential linearized augmented planewave
(LAPW) method, with the augmented planewave plus local orbital implementation.
\cite{wien,sjostedt,singh_book}
LAPW spheres of radii 2.3$a_{\rm 0}$ for Ba, Ir and
2.1$a_{\rm 0}$ for Fe, As, Ru were employed.
We used well converged basis sets determined
by $R_{\rm min}K_{\rm max}$=8.5, where $R_{\rm min}$ is the radius
of the smallest LAPW sphere and $K_{\rm max}$ corresponds to the
planewave cutoff for the interstitial region.
We included relativity at the scalar relativistic level.
Local orbitals were included to accurately treat the semi-core states.
The Brillouin zone sampling for self-consistent calculations was
done using the special \textbf{k}-point method,
with a 24x24x24 grid for body-centered ThCr$_{2}$Si$_{2}$
structure and a 21x21x9 grid for the quadrupled
tetragonal supercells with partial Ru and Ir
substitution (see below).

\section{Ruthenium Substitution}

\begin{figure}
\includegraphics[width=2.5in,angle=0]{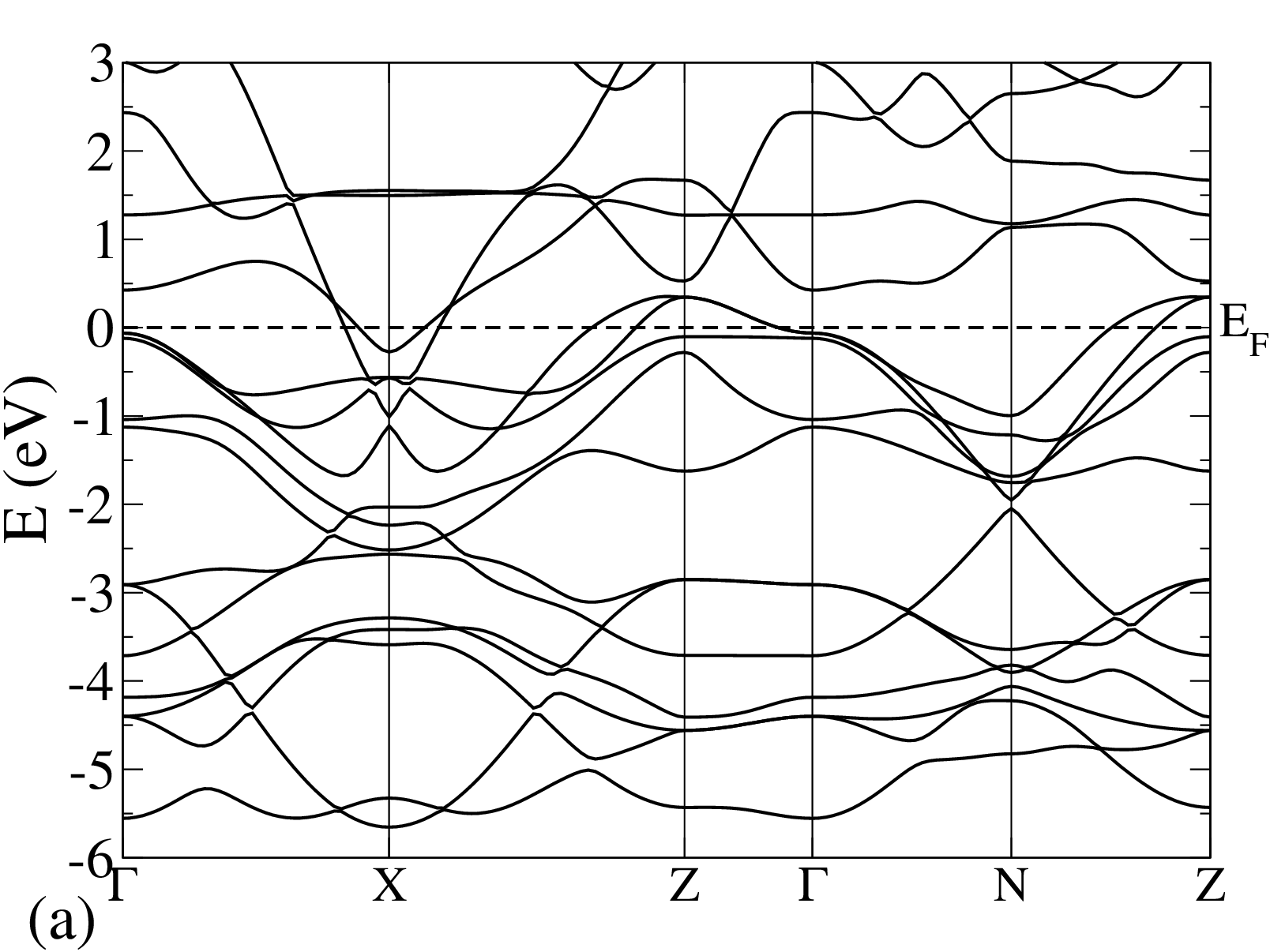}
\includegraphics[width=2.5in,angle=0]{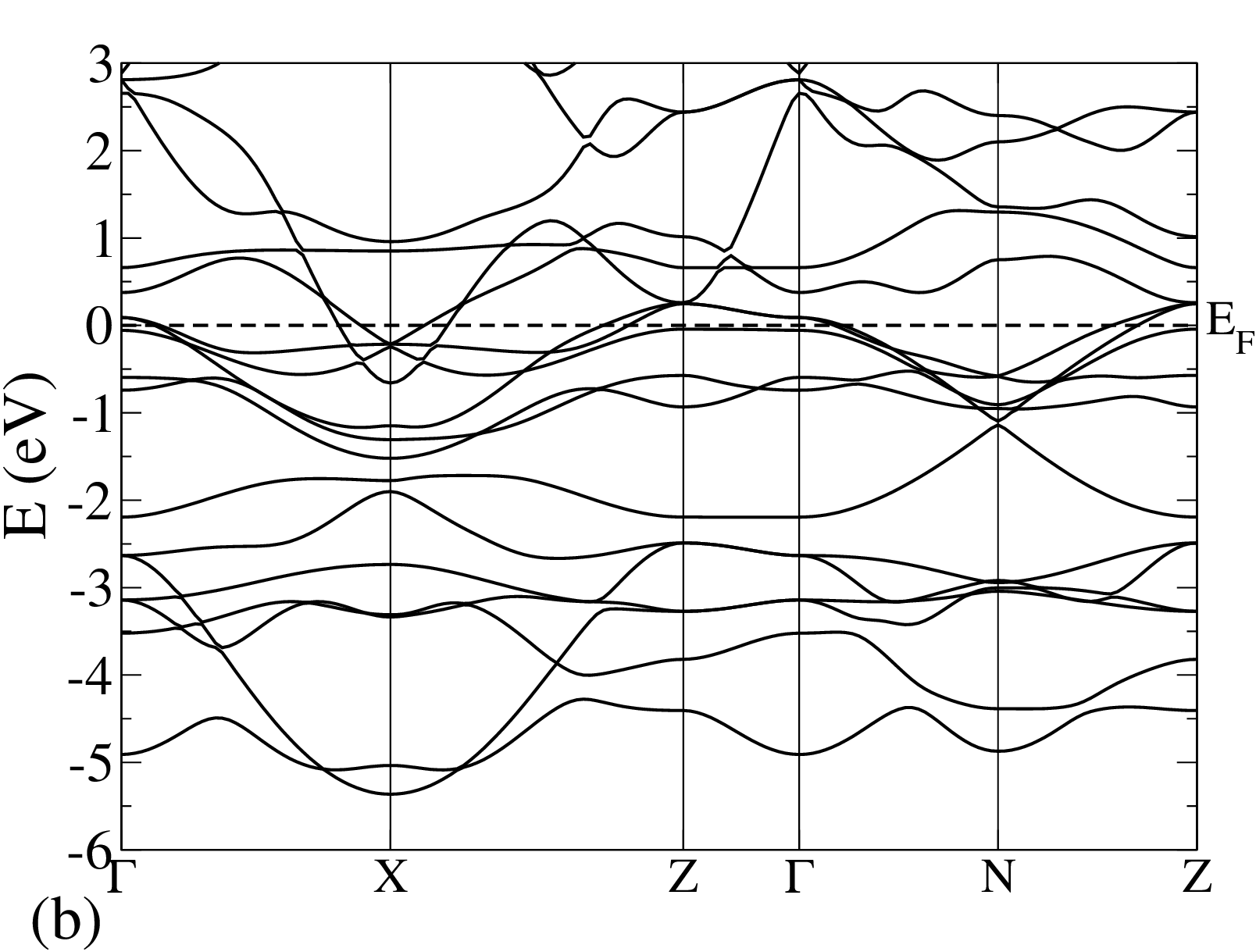}
\caption{Calculated electronic band structure of
(a) BaRu$_{2}$As$_{2}$ with the relaxed $z_{\rm As}$,
comparing with that of
(b) BaFe$_{2}$As$_{2}$.
The bands are plotted along high-symmetry directions in the
body-centered tetragonal Brillouin zone, and $X$ point corresponds to the zone
corner $M$ point in the tetragonal Brillouin zone.
For BaFe$_{2}$As$_{2}$,
we used the experimental lattice parameter
$a$ = 3.9625 {\AA} and $c$ = 13.0168 {\AA} (Ref. \onlinecite{fe_latt})
and non-spin-polarized GGA optimized As height $z_{\rm As}$ = 0.344.}
\label{bands}
\end{figure}

We first calculated the electronic structure of completely Ru substituted
non-superconducting BaRu$_{2}$As$_{2}$ to check
whether additional bands or significant changes in the
band shapes are introduced by Ru.
The experimental tetragonal lattice constant
$a$ = 4.152 {\AA} and $c$ = 12.238 {\AA}\cite{ru_latt}
were employed and the internal As coordinate $z_{\rm As}$ were optimized
by total energy minimization as 0.3528.
For this material, we checked for but did not find
magnetic order of either
ferromagnetic or antiferromagnetic character,
consistent with recent experiments.\cite{baru2as2}
Fig. \ref{bands} shows the calculated band structure,
in comparison with that of
non-spin-polarized BaFe$_{2}$As$_{2}$ as obtained with the same approach.
As may be seen, the general features of band structures around
$E_{F}$ for two compounds show quite close similarity and no new bands appear
in BaRu$_{2}$As$_{2}$.
In particular, there are similar compensating electron Fermi surface
sections at the zone corner and hole sections at the zone
center ($\Gamma$-Z direction).
Especially at the zone corner, there is a very strong similarity of the
electron cylinders in the two compounds.
However, for BaRu$_{2}$As$_{2}$ we do find a somewhat different structure
of the hole Fermi surfaces, which are
centered near Z as compared to BaFe$_{2}$As$_{2}$ where
hole bands exist along the $\Gamma$-Z direction.
This yields a closed hole Fermi surface in the $k_z$ direction
in BaRu$_2$As$_2$.
This more 3D shape is suggestive of more Ru -- As hybridization,
which is also seen in the projections of the electronic density of 
states (DOS).
In any case,
the position of $E_F$ with respect to the compensating position between
the hole and electron band edges is maintained similar to non-spin-polarized
BaFe$_2$As$_2$.
This is different from the non-isoelectronic compounds,
BaCo$_{2}$As$_2$\cite{baco2as2} and
BaNi$_{2}$As$_2$,\cite{bani2as2} which also show rather
similar electronic structures to the Fe compound,
but with the $E_{F}$ shifted to higher energy corresonding to the
higher electron count. This shift of the Fermi energy results in
dramatically different physical properties.

\begin{figure}
\includegraphics[width=3.0in]{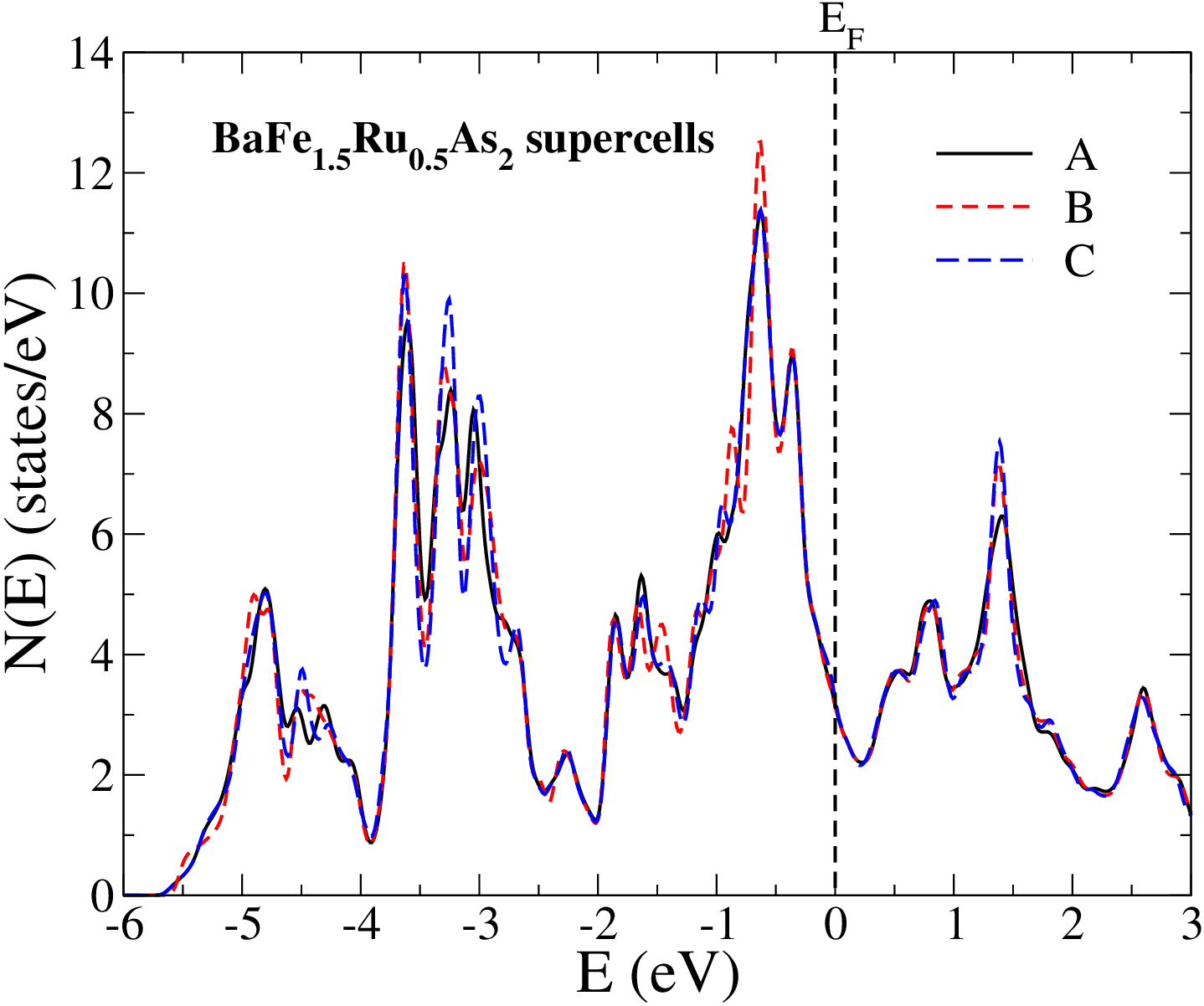}
\caption{(Color online)
Comparison of the calculated electronic DOS for the three
BaFe$_{1.5}$Ru$_{0.5}$As$_{2}$ supercells, $A$, $B$ and $C$
as described in the text.
}
\label{ss-comp}
\end{figure}

\begin{figure}
\includegraphics[width=3.0in]{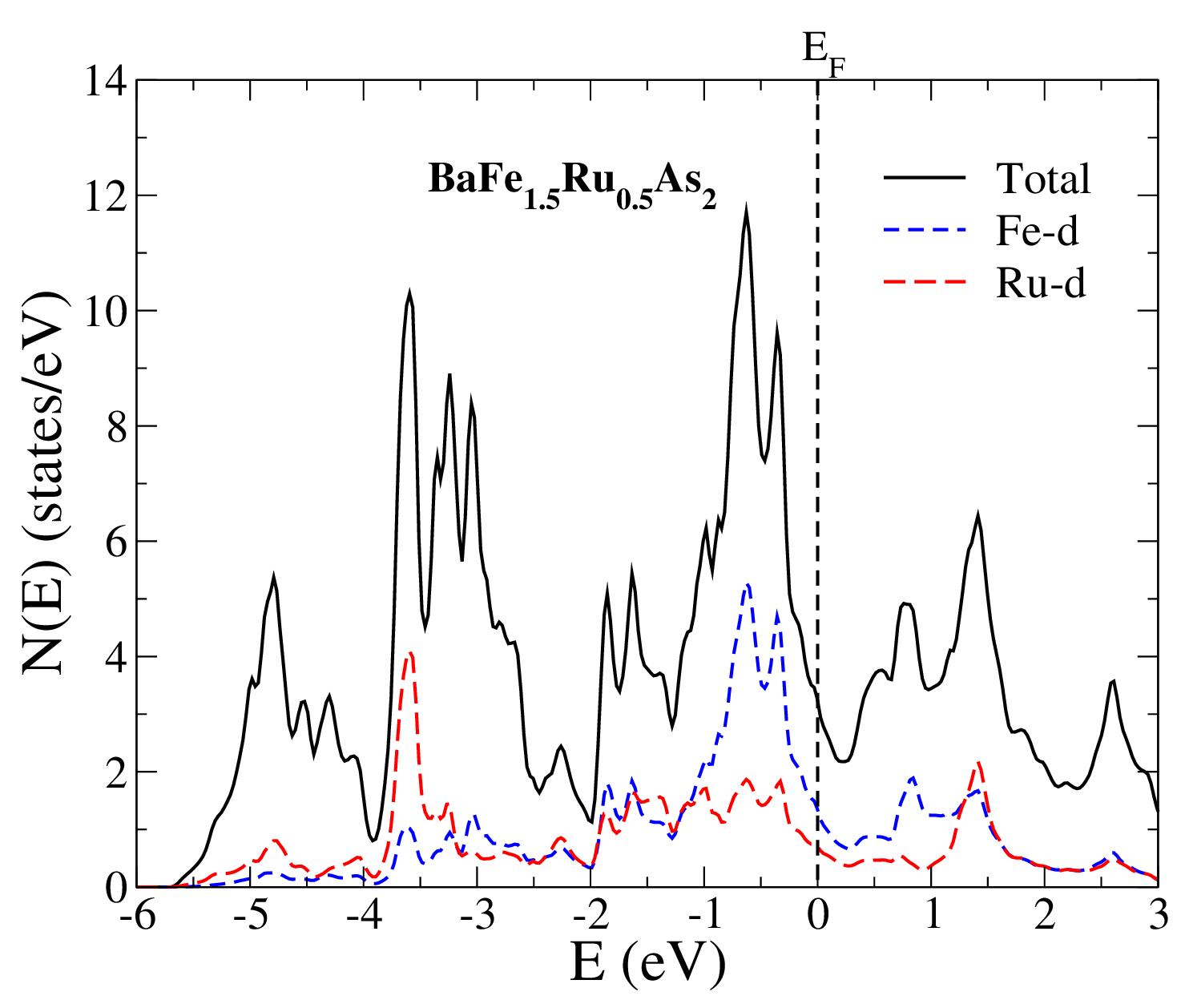}
\includegraphics[width=3.0in]{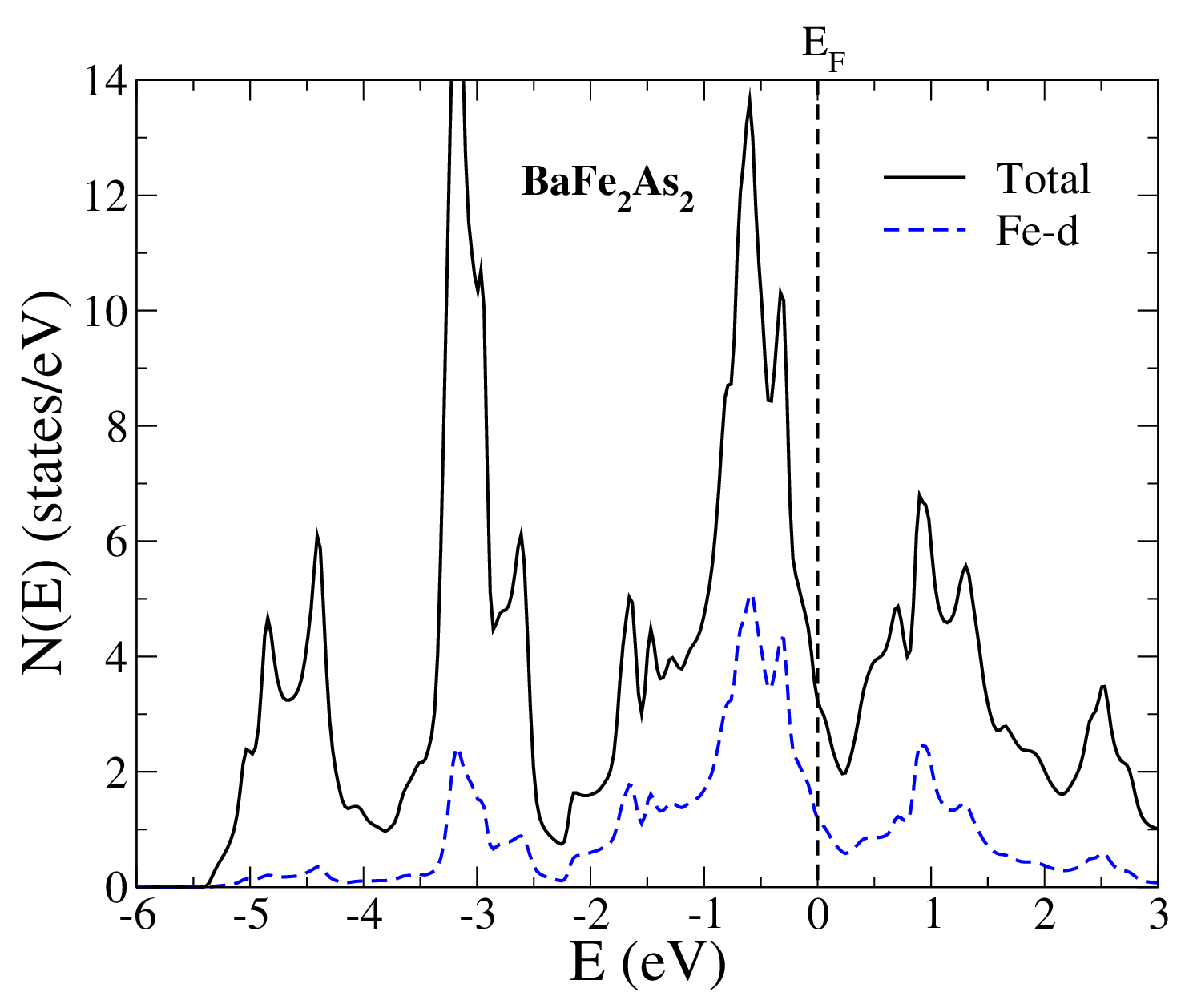}
\caption{(Color online)
Calculated total DOS and projections
for BaFe$_{1.5}$Ru$_{0.5}$As$_{2}$ (top),
using the quadrupled supercell $A$ (see text).
The total DOS is on a per formula (BaFe$_{1.5}$Ru$_{0.5}$As$_{2}$) basis
while the projections are per atom.
The total and Fe $d$-projected DOS for BaFe$_{2}$As$_{2}$ (bottom)
are presented for comparison.
}
\label{Ru-Fe_dos}
\end{figure}

\begin{figure}
\includegraphics[width=3.2in]{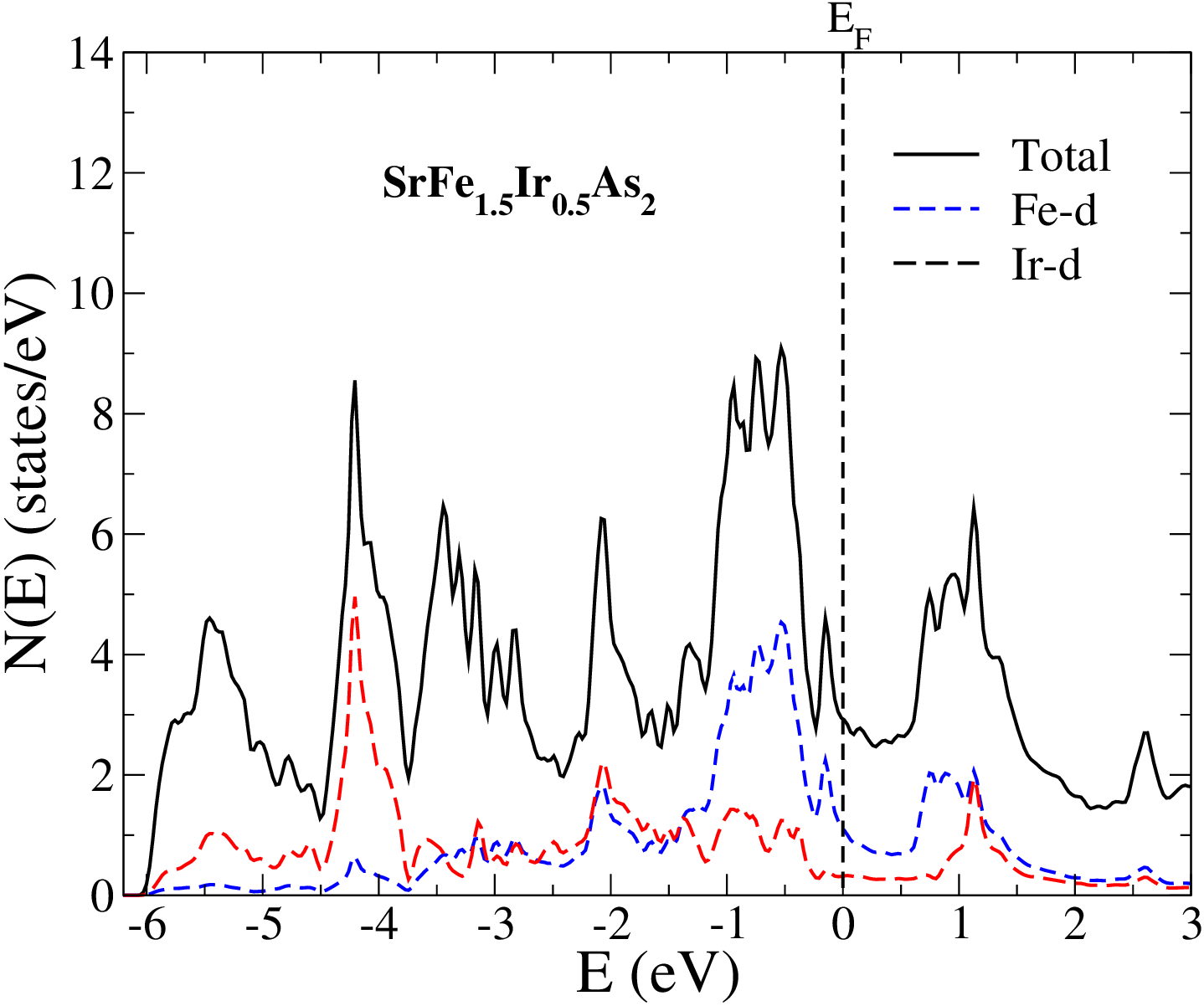}
\caption{(Color online)
Total and projected electronic DOS of
SrFe$_{1.5}$Ir$_{0.5}$As$_{2}$,
calculated in the same way as that of BaFe$_{1.5}$Ru$_{0.5}$As$_{2}$.
The tetragonal lattice parameters from experiment (Ref. \onlinecite{sr_ir})
$a$ = 3.95 {\AA} and $c$ = 12.22 {\AA} were used for supercell simulation.}
\label{Ir_dos}
\end{figure}

We used supercell calculations to explicitly study the effects of
partial Ru and Ir substitution.
For this purpose we used a $\sqrt{2}$x$\sqrt{2}$ doubling
in plane, and also doubled the cell along the $c$-axis by going to
the conventional tetragonal cell as opposed to the body centered cell.
This leads to a quadrupling of the ThCr$_2$Si$_2$ structure cell.
We then replaced one Fe by Ru or Ir in each plane.
This corresponds to a composition of BaFe$_{1.5}$Ru$_{0.5}$As$_{2}$
(nominal Ba$_{4}$Fe$_{6}$Ru$_{2}$As$_{8}$) or
SrFe$_{1.5}$Ir$_{0.5}$As$_2$.
For Ba$_{4}$Fe$_{6}$Ru$_{2}$As$_{8}$ we did calculations for three
different arrangements of the Ru. These were with the two Ru directly
on top of each other ($A$), shifted by one Fe-Fe distance ($B$) and
shifted by one lattice parameter, $a$ ($C$).
The tetragonal lattice parameter for BaFe$_{1.5}$Ru$_{0.5}$As$_{2}$,
$a$ = 3.98 {\AA} and $c$ = 12.95 {\AA} were taken from Ref. \onlinecite{ru_1}
and the internal coordinates were optimized by energy minimization.
As shown in Fig. \ref{ss-comp}, the electronic structures for these
three supercells are very similar. The values at the Fermi energy for the
three cells are
$N(E_F)$=3.17 ($A$), $N(E_F)$=3.14 ($B$), and $N(E_F)$=3.17 ($C$).
In the following we focus on results for supercell $A$, since
the same conclusions would be drawn based on the others.

It is found that, as the result of larger size for Ru$^{2+}$,
apart from expanding distances within Fe-Ru planes,\cite{ru_1}
the Ru-As distance (2.41 {\AA}) is also slightly larger than the
Fe-As distances (2.30 {\AA} -- 2.35 {\AA}), and accordingly the Ba layers
are also distorted.
The calculated DOS is shown in
Fig. \ref{Ru-Fe_dos} in comparison with that of BaFe$_{2}$As$_{2}$.
We did not find any additional peak associated with extra flat bands
introduced by Ru substitution.
This and the shape of the DOS reflecting bands of mixed Fe and Ru character,
as opposed to separate Fe and Ru derived bands,
indicate that the electronic structure of
the Fe-Ru system is that of a coherent alloy as might be anticipated
from the similar electronic structures of
BaRu$_{2}$As$_{2}$ and BaFe$_{2}$As$_{2}$.
In addition, it can be clearly seen that the shape of total
DOS near $E_{F}$ is almost unchanged after Ru substitution,
except that the values decrease somewhat.
This decrease
mainly results from the enlarged $d$ band width as the result of
increased hopping between transition metals and hybridization
involving Ru.
$N(E_{F})$ is reduced to 
3.17 eV$^{-1}$ per BaFe$_{1.5}$Ru$_{0.5}$As$_{2}$,
as compared to 3.28 eV$^{-1}$ for BaFe$_{2}$As$_{2}$
(note that these GGA values are somewhat larger than the LDA values,
e.g. $N(E_F)$=3.06 eV$^{-1}$ for BaFe$_2$As$_2$
in Ref. \onlinecite{bafe2as2}).
In any case, we can conclude that no additional electron carriers are
introduced to Ru alloyed system in the sense
that the band filling is unchanged.

The general shape of our density of states is similar to that reported by
Paulraj and co-workers. \cite{ru_1}
However, there are significant differences in detail.
As mentioned, we
find a decrease in the value of $N(E_F)$ upon alloying rather than
an increase.
We find that for BaFe$_{1.5}$Ru$_{0.5}$As$_{2}$
the distribution of Fe $d$ states is almost the
same as in non-spin-polarized BaFe$_{2}$As$_{2}$,
giving the main contribution to the states near $E_{F}$ and
only modestly mixing with As $p$ states.
In contrast, Ru $d$ derived bands form a peak at $\sim$-3.5 eV and overlap
in energy with As $p$ states concentrated within the region of
-5.5 eV to -2.5 eV, and thus more strongly hybridized with them in
spite of the larger Ru-As distance.
This results in a reduced Ru $d$ contribution to the states
near $E_{F}$.
Thus the contribution to $N(E_{F})$ from Ru $d$ states is lower by
half than that of the Fe $d$ states. This plays an important role in
the
decreased $N(E_{F})$ for the BaFe$_{1-x}$Ru$_{x}$As$_{2}$ system
and is a consequence of Ru-As hybridization.

\section{suppression of the SDW}

As mentioned, understanding the suppression of the SDW order is
important.
It is convenient to discuss the magnetic susceptibility,
$\chi({\bf q})$, which is given in the random-phase approximation
by the enhanced Lindhard susceptibility
$\chi({\bf q})=\chi_0({\bf q})~[1-I({\bf q})\chi_0({\bf q})]^{-1}$,
where 
$\chi_0({\bf q})$ is the bare susceptibility, which is a density of
states like term
and $I({\bf q})$ is the Stoner parameter, whose ${\bf q}$ dependence
reflects the changing band character on the Fermi surface.
In BaFe$_2$As$_2$ and other undoped Fe-based materials,
the high $N(E_F) \propto \chi_0(0)$
with the large Stoner parameter $I$ ($\sim$0.9 eV,
characteristic of a 3$d$ transition element,
with $N(E_F)$ now on a per spin per Fe basis)
\cite{andersen} 
puts them near the Stoner criterion ($N(E_{F})I >$ 1) for itinerant
ferromagnetism magnetism. \cite{singh-du}
Further, the strong Fermi nesting between approximately
size-matched electron and hole sections yields on top of this a
peak in $\chi_0({\bf q})$ at the zone corner.
This explains the SDW instability at the nesting
vector (1/2,1/2).

Turning to the Ru alloyed system, on the one hand,
more hybridization with As $p$ states and the larger size of the Ru
4$d$ orbital relative to the
Fe 3$d$ orbital cause a reduction in $N(E_{F})$, as mentioned.
On the other hand, the atomic-like $I$ is reduced to the value
of $\sim$0.6 eV for the mid-4$d$ transition metal, Ru.\cite{andersen}
For this coherent alloy system,
the Stoner parameter is the average of the atomic Stoner parameters
weighted by the squares of the partial contributions of the
different $d$ orbitals to the $\chi_0$ (i.e. $N(E_F)$ for ${\bf q}$=0).
\cite{andersen,averageI}
The rather smaller $N(E_{F})$ of Ru $d$ states as mentioned would further
reduce the contribution from Ru to the average Stoner parameter.
Considering both $I$ and $N(E_F)$ it is clear that the
Ru alloy is much less magnetic than the pure Fe compound.
The same factors will apply to the SDW instability, and provide
the explanation for its suppression upon Ru alloying seen both
experimentally, and in our direct density functional calculations.

\section{iridium substitution}

Finally,
we also calculated the electronic structure for 5$d$ transition metal,
Ir alloyed system, SrFe$_{1.5}$Ir$_{0.5}$As$_{2}$,
As shown in Fig. \ref{Ir_dos}, the substitution of Fe with Ir also leads
to an electronic structure characteristic of a coherent alloy.
Compared with BaFe$_{1.5}$Ru$_{0.5}$As$_{2}$,
the scale of total DOS is further reduced,
reflecting still stronger hybridization and expanded band width.
This is seen for example in the large Ir contribution to the 
As $p$ derived bands at $\sim$ -4 eV and the correspondingly
reduced Ir contribution to the metal $d$ bands from
$\sim$ -2 eV to 2 eV relative to $E_F$.
More importantly, $E_{F}$ clearly shifts upwards towards the
bottom of the pseudogap, which reflects additional electrons that are
doped into the $d$ bands
by alloying this Co column transition metal.
The doping amounts to one carrier per Ir since, as for Ru, the electronic 
structure is coherent and no additional bands are introduced.
The value of $N(E_{F})$ is 2.91 eV$^{-1}$ per formula of
SrFe$_{1.5}$Ir$_{0.5}$As$_{2}$.
In addition, we note that
the Stoner parameter of Ir is rather low, $I_{\rm Ir}$ $\sim$ 0.5 eV.
\cite{andersen}
Thus, the suppression of SDW order in this Ir-alloyed system\cite{sr_ir}
may be attributed to effects of both a
decreased Stoner factor and electron doping.

\section{summary and conclusions}

In summary,
we show based on density functional calculations
the alloying BaFe$_2$As$_2$ with Ru and SrFe$_2$As$_2$ with Ir
results in the formation of a coherent electronic structure without
additional localized states associated with the impurity atoms.
As such,
the suppression of SDW magnetic order
for the Ru alloyed Ba(Sr)Fe$_{2}$As$_{2}$ system is due to the increased
hybridization and a reduced average Stoner parameter reflecting the
different chemistry of $3d$ and $4d$ transition elements.
A similar consideration applies to alloying with Ir, although in that
system Ir also provides doping.
We note that scattering due to Ru/Ir disorder in the actual alloys
might also play a role in suppressing the SDW, since scattering
generally works against a nesting induced instability.
Also it
should be mentioned that the detailed role of 4$d$ and 5$d$ transition
metals on the interplay
between magnetism and superconductivity in Fe-based superconductors
might be quite complex.
For example, Ru alloying is found to completely suppress
SDW order in the PrFe$_{1-x}$Ru$_{x}$AsO system but no superconductivity
emerges, at least in samples investigated to date.\cite{1111_ru}
However,
Ir doped LaFe$_{1-x}$Ir$_{x}$AsO is reported to show
superconductivity with $T_{c}$ up to 10.5 K.\cite{1111_ir}
Further studies will be helpful in understanding these differences.
In any case, substitution of Fe with 4$d$ and 5$d$ transition metals
in Fe-based materials provides another avenue for suppressing SDW
magnetic order and thereby inducing superconductivity in Fe-based materials,
even in the absence of doping.

\acknowledgements

We are grateful for helpful discussions and
assistance from A. Subedi.
This work was supported by the Department of Energy,
Division of Materials Sciences and Engineering.

\bibliography{BaFeRuAs2}
\end{document}